\documentclass[prb,twocolumn,floatfix,showkeys]{revtex4-1} %,preprint
\usepackage{amsmath,amssymb}
\usepackage{graphicx} 
\usepackage{epstopdf}
\usepackage{color} 

\def\rmd{{\rm d}}
\def\sgn{{\rm sgn}}
\def\csch{{\rm csch}}
\usepackage{subfigure}
\usepackage[colorlinks,allcolors=blue]{hyperref}
\def\x{\end{document}}

\makeatletter
\newcommand*{\defeq}{\mathrel{\rlap{%
                     \raisebox{0.3ex}{$\m@th\cdot$}}%
                     \raisebox{-0.3ex}{$\m@th\cdot$}}%
                     =}
\makeatother

\begin{document}

\title{Bound states of a two-dimensional electron gas in inhomogeneous 
magnetic fields} 

\author{Siddhant Das} 
\email {siddhantdas@yahoo.co.in} 
\affiliation{Electronics and Communication Engineering, 
National Institute of Technology, Tiruchirappalli 620015, TN, India}

\date{August 15, 2015}

\begin{abstract}
We study the bound states of a two dimensional free electron gas (2DEG) subjected to a perpendicular inhomogeneous magnetic field. An analytical transfer matrix (ATM) based exact quantization formula is derived for magnetic fields that vary (arbitrarily) along one spatial direction. As illustrative examples, we consider (1) a class of symmetric power law magnetic fields confined within a strip, followed by the problem of a (2) 2DEG placed under a thin ferromagnetic film, which are hitherto unexplored. The exact Landau levels for either cases are obtained. Also, the role of the fringing magnetic field (present in the second example) on these levels is discussed.
\end{abstract}


\maketitle

\section{Introduction} \label{sec1}

The quantum mechanics of electrons constrained in two dimensions has gained 
a lot of interest since the advent of the (Integer) Quantum Hall and 
Fractional Quantum Hall Effects in the 1980s. Over the years, the 
experimental techniques of probing these systems with precisely structured 
magnetic fields at low temperatures ($\approx 100$ mK) have been perfected. 
Presently, magnetic fields that vary appreciably (even) in the nanometer 
scale can be created by fabricating thin films of (a) metal-excitable with a 
calculated current distribution, (b) ferromagnetic materials, (c) type-I 
superconducting materials, on a two-dimensional electron gas (2DEG) system. 
Interestingly, the magnetic field in these cases can be described in closed 
analytic form\cite{1,2} which, is an invitation for making very precise 
predictions for the behavior of 2DEGs subjected to such inhomogeneous 
fields. However, on the theoretical side we face a setback in considering 
arbitrary magnetic fields, as the Schr\"{o}dinger equation describing the 
electron-field interaction can seldom be solved analytically, except in very 
simple cases. Often, powerful numerical methods offer a solution to this 
problem, albeit, at the loss of significant physical insight. 

Considering the difficulty of a generic 2DEG-magnetic field interaction 
problem, we focus on a select class of magnetic fields (perpendicular to the 
plane of the electron gas) that are confined within an infinitely long strip 
of width $d$. Further, if the magnetic field varies across the strip 
(only), remaining translation invariant along the length of the strip, an 
exact enumeration of the bound state energies and tunneling probabilities of 
the electron is possible for any form of inhomogeneity. We focus on the 
bound state problem here, reserving a discussion of scattering for a later 
paper.

Besides, a significant chunk of the literature on magnetic strips is devoted 
to a study of electron tunneling through the magnetic field barrier\cite{2,3,4,5} while, relatively little is explored on the bound state solutions. Bound state solutions of a linearly varying magnetic field were obtained by 
M\"{u}ller.\cite{3} Even in this simple case, it is not possible to describe 
the 
electron wave functions in terms of any special function or finite analytic 
combinations thereof. Even otherwise, bound state solutions are exceedingly 
special, as their existence is not necessarily guaranteed for a given 
magnetic field while, scattering states always exist (for any given field 
variation) when the energy is above a minimum threshold value. Another 
nontrivial field variation that enjoys exact solvability is $\mathbf{B} 
=1-\tanh^2x~\mathbf{e}_{z}$\cite{5} in which, the discrete and 
continuous part of the spectrum overlap in an energy range and are selected 
by the $y$ momentum associated with the wave function.

The main goal of this paper is to find the exact bound state energies 
(Landau levels) for any given magnetic field variation (whenever such levels exist). We formulate the problem in Section~\ref{sec2} obtaining an effective 
one-dimensional magnetic potential for the electron. The criterion for the 
allowed bound state energies is obtained with an analytic transfer matrix 
(ATM) approach in Section~\ref{sec2.1}. Section~\ref{sec3} is devoted to 
examples. Firstly, in Section~\ref{sec3.1} we obtain the Landau levels (LLs) 
of a magnetic strip that has a symmetric power law field variation. Following 
which, we consider the problem of a 2DEG placed under a ferromagnetic film in 
Section~\ref{sec3.2}. Unlike in the former example, the magnetic field in this 
case offers a fringing field outside the strip which, has a significant 
effect on the LLs. We conclude in Section~\ref{sec4} outlining avenues of 
further study. An appendix at the end gives the proof of an important result 
used in Section~\ref{sec2.1}.

\section{Problem formulation}\label{sec2}

We place the 2DEG on the $x$--$y$ plane, subjected to a perpendicular magnetic 
field
\begin{equation}\label{eq1}
\mathbf{B} = B_{o}  B(2x/d) \mathbf{e}_{z}, \quad
\zeta \mapsto B(\zeta)\ne (=)0,\quad |\zeta|\le (>)1  
\end{equation}
where, $B_{o}$ is the field strength and $d$ is the width of the strip. 
$B(\zeta)$ must be integrable. The vector potential for this 
field, in the Landau gauge reads
\begin{equation}\label{eq2}
\mathbf{A} =\frac{B_{o} d} {2} \Phi (2x/d) \mathbf{e}_{y}, 
\quad
\Phi (\zeta)\defeq\int_{-\infty}^\zeta {B(\zeta')\rmd  \zeta'}. 
\end{equation}
We set up a minimal coupling Hamiltonian ${H} =({p} +e\mathbf{A})^{2}/2m^{\ast
}$ 
to describe the electron-field interaction where, $m^{\ast}$ is the effective 
mass of the electron with charge $-e$. The magnetic length $\ell_{B} 
=\sqrt {\hslash /eB_{o}}$ and cyclotron frequency $\omega_{c} =eB_{o}/m^{\ast
}$ 
provide natural length and time scales in this problem. Scaling the energy $E
\mapsto (\hslash \omega_{c} /2) \varepsilon$ and the coordinates 
$(x,y)\mapsto \ell_{B} (\xi,\eta)$ we obtain the 
Schr\"{o}dinger equation 
\begin{widetext}
\begin{equation}\label{eq3}
\left\{{\nabla}_{\xi \eta}^{2} +i\frac{d} {\ell_{B}} \Phi 
\left(\frac{2\ell_{B}}{d} \xi \right)
\partial_{\eta}-\left(\frac{d} {2\ell_{B}} 
\Phi \left(\frac{2\ell_{B}} {d} \xi \right) \right)^{2}\right\} 
\psi \left(\xi,\eta \right)
=-\varepsilon \psi \left(\xi,\eta \right) 
\end{equation}
\end{widetext}
satisfied by the wave function $\psi(\xi,\eta)$. The width of 
the strip in $\ell_{B}$ units is given by $2\beta~(\beta\defeq d/2\ell_{B})$. 
Since, the Hamiltonian has a translation symmetry in the $y$ direction, the 
commutation identity $[{H},{p}_{y}]=0$ holds 
good. Thus, $\psi$ is a simultaneous eigenstate of ${H}$ and 
${p}_{y}$. An ansatz $\psi (\xi,\eta)=e^{iq\eta} \varphi 
(\xi)$ would satisfy this requirement provided, $\varphi$ 
solves the one-dimensional Schr\"{o}dinger equation
\begin{equation}\label{eq4}
\frac{\rmd^{2} \varphi} {\rmd\xi ^{2}} 
+\left(\varepsilon-V_{q} (\xi) \right)\varphi =0,\quad 
{V_{q} (\xi)\defeq\left(q+\beta \Phi (\xi /\beta) \right)}^{2}. 
\end{equation}
We call $V_{q} (\xi)$ the effective magnetic potential whose, 
shape is modulated by the $y$ momentum $(\hslash/\ell_B) q$ associated with 
the wave function. This makes the quantum mechanical behavior wave-vector 
dependent.\cite{2} We will explore many interesting possibilities that arise 
(concerning the existence of bound states) due to the presence of $q$. 
Note that outside the magnetic strip i.e. $|\xi|>\beta,~V_{q} (\xi)$ is constant as shown in Fig.~\ref{fig1}. Specifically, 
for $\xi <-\beta,~V_{q} (\xi)=q^{2} $, while for $\xi 
>\beta, V_{q} (\xi)=(q+\beta \Phi (1))^{2} $ where, $\Phi (1)$ is 
proportional to the magnetic flux per unit length linked with the infinite 
strip (see Equation~\eqref{eq2}). Although, bound state solutions of equation~
\eqref{eq4} can be anticipated for energies $\varepsilon <\min \left\{q^{2},
(q+\beta \Phi (1))^{2}\right\}$; for a given $B(\zeta)$, the effective 
potential may not offer `wells' for containing the electron, in which case 
bound state solutions will not exist for any $\varepsilon$. This shape 
dependence makes bound states rather scarce unlike scattering solutions. 

\subsection{Formally exact criterion for landau levels} \label{sec2.1}

In this section we use the analytic transfer matrix method (ATMM) to obtain 
an exact criterion for the allowed bound states in the effective 
potential $V_{q} (\xi)$. The ATMM emerged in the problem of 
finding guided modes of the electromagnetic field in a graded-index optical 
fiber.\cite{6,7} The method was readily extended by the pioneers to apply to 
one-dimensional problems in quantum mechanics\cite{8,9,10,11,12} where, it has been remarkably successful not only as a calculation device for exact energy levels but also as a conceptual tool; providing deeper insights into the working and limitations of semi classical quantization schemes like the 
Bohr--Sommerfeld and the WKB method (and refinements of the same).\cite{10}
These efforts also led to the conceptualization of the `modified momentum' 
which, substituted for the canonical momentum makes the Bohr-Sommerfeld 
quantization exact.\cite{9} The ATMM, combined with super-symmetric techniques has also been applied to nontrivial potentials yielding promising results.\cite{13} 

In order to keep the paper self-contained, we give a complete derivation of 
the ATM quantization condition clarifying, a major ingredient in the 
derivation (the phase losses at the classical turning points) which, in our 
opinion has not been rigorously justified in previous accounts (Ref. appendix). Additionally, we evaluate the ATM quantization integral in closed form which, is a new development. 

\begin{figure}
\centering
\includegraphics[width=\columnwidth]{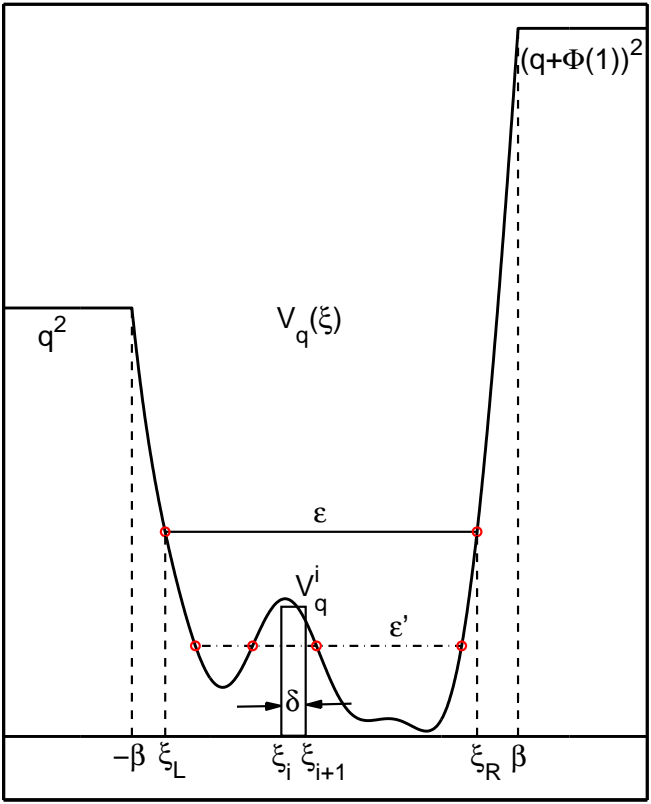}
\caption{Sketch of the effective magnetic potential $V_{q} (\xi)$ 
with classical turning points $\xi_{L,R}({\color{red}\boldsymbol{\circ}})$ for 
energy $\varepsilon$.\label{fig1}}
\end{figure}

Consider two classical turning points $\xi_{L} $ and $\xi_{R} $ that 
solve $V_{q} (\xi)=\varepsilon$ in the region $(-\beta,\beta)$. The case of 
more than two turning points (as with the energy $\varepsilon'$ in 
Fig.~\ref{fig1}) is not addressed in this paper, as the ATMM is not easily 
generalized for multiple turning points. We partition the 
sub intervals $(-\beta,\xi_{L}), (\xi_{L},\xi_{R})$ and 
$(\xi_{R},\beta)$ into $l, m$ and $n$ 
segments respectively, each with width $\delta$. Thus, any intermediate 
point $\xi_{i} =i\delta-\beta,~i=0,1,2\ldots (l+m+n)$ with $\xi_{l+m+n} =\beta$. 
Certainly, 
$\xi_{L} =l\delta-\beta$ and $\xi_{R} =(l+m)\delta-\beta$. The continuous 
potential is now replaced by a piecewise constant equivalent over these segments, such that the potential in the segment $(\xi_{i},\xi_{i+1})$ is
$V_{q}^{i}\defeq V_{q} \left(\frac{\xi_{i} +\xi_{i+1}} {2}\right)$. Further, 
the solution of equation~\eqref{eq4} in this segment is given by
\begin{equation}\label{eq5}
\varphi (\xi)=A_{i} e^{i\kappa_{i} (\xi-\xi_{i+1})} +B_{i} e^{-i\kappa_{i} (\xi-\xi_{i+1})},\quad
 \kappa_{i}=\sqrt {\varepsilon-V_{q}^{i}}  
\end{equation}
where, $A_{i} (B_{i})$ is the probability amplitude for the forward 
(backward) traveling wave component. In equation~\eqref{eq5} $\varphi$ may be 
tagged explicitly to show the correspondence with the $i$\textsuperscript{th} segment. We prefer to infer this from the context. Necessitating the continuity of the wave function and its derivative at the endpoints of the segment we arrive at the matrix equation
\begin{equation}\label{eq6}
\begin{bmatrix}
\varphi (\xi_{i})\\
\dot{\varphi} (\xi_{i})\\
\end{bmatrix}=M_{i} 
\begin{bmatrix}
\varphi (\xi_{i+1})\\
\dot{\varphi} (\xi_{i+1})\\
\end{bmatrix},~M_{i}=
\begin{bmatrix}
\cos (\kappa_{i} \delta) &-\frac{\sin (\kappa_{i} \delta)}{\kappa_{i}}
\\
\kappa_{i} \sin (\kappa_{i} \delta) & 
\cos (\kappa_{i} \delta)\\
\end{bmatrix} 
\end{equation}
where, the overhead dot denotes differentiation w.r.t $\xi$. Note that 
$\kappa_{i} \ne 0$, since we assume no more than two classical turning points at this stage. Few authors prefer separate formulae for the transfer matrix 
$M_{i}$ that hold when $\varepsilon >V_{q}^{i}$ or otherwise. In view of 
this distinction, our expression corresponds to the former case, while the 
latter case i.e. $\varepsilon <V_{q}^{i}$, leading to an imaginary $\kappa 
_{i}$ is easily addressed with the analytic continuation of the 
trigonometric functions into the complex plane in equation~\eqref{eq6}. Thus, 
recovering the other expression for the transfer matrix when the case 
arises. With this caveat we overcome the need for selective indexing of 
transfer matrix products.

Left multiplying equation~\eqref{eq6} with $\begin{bmatrix}
-\dot{\varphi} (\xi_{i}) & \varphi (\xi_{i}) 
\end{bmatrix}$ 
and dividing by $\varphi (\xi_{i})\varphi (\xi_{i+1})$, 
we obtain 
\begin{equation}\label{eq7}
\begin{aligned}
&\begin{bmatrix}P_{i}  & 1\end{bmatrix}M_{i}
\begin{bmatrix}1\\-P_{i+1} \\\end{bmatrix}=0, 
\qquad
P_{i}\defeq-\frac{\dot{\varphi}(\xi_{i})}{\varphi (\xi_{i})}\\
&\Rightarrow\frac{P_{i}} {\kappa_{i}}=\frac{\frac{P_{i+1}} {\kappa_{i}}
-\tan (\kappa_{i} \delta)} {1+\frac{P_{i+1}} {\kappa_{i}} 
\tan (\kappa_{i} \delta)}. 
\end{aligned}
\end{equation}
Through the tangent addition formula we obtain
\begin{gather}\label{eq8}
%\begin{aligned}
\frac{P_{i}} {\kappa_{i}} =\tan \left[\tan^{-1} 
\left(\frac{P_{i+1}} {\kappa_{i}} \right)-\kappa_{i} \delta\right] \nonumber\\
\Rightarrow \tan^{-1} \left(\frac{P_{i}} {\kappa_{i}} \right)-\tan^{-1} 
\left(\frac{P_{i+1}} {\kappa_{i}} \right)
=z_{i} \pi-\kappa_{i} \delta,\,\nonumber\\z_{i} =0, 1, 2,\ldots
%\end{aligned}
\end{gather}
Note that $P(\xi)$ satisfies the Riccati equation 
\begin{equation}\label{eq9}
\dot{P} =P^{2} +\varepsilon-V_{q} (\xi)=P^{2} +\kappa^{2}  
\end{equation}
which, has intimate connections with the Schr\"{o}dinger equation~\cite{14}. 
It is often advantageous (as in the present case) to develop identities 
involving $P$. Rearranging equation~\eqref{eq8} and summing over 
$i$ from $l+1$ to $l+m$ yields
\begin{equation}\label{eq10}
\begin{aligned}
&\sum\limits_{i=l+1}^{l+m}  {\kappa_{i} \delta}  +\sum\limits_{i=l+1}^{l+m}  
\left[\tan^{-1} \left(\frac{P_{i+1}} {\kappa_{i+1}} \right)-\tan 
^{-1} \left(\frac{P_{i+1}} {\kappa_{i}} \right) \right]\\
&\quad =N\pi 
+\sum\limits_{i=l+1}^{l+m}  \left[\tan^{-1} \left(\frac{P_{i+1}} {\kappa 
_{i+1}}\right)-\tan^{-1} \left(\frac{P_{i}} {\kappa_{i}} \right) \right]\\ 
&\Rightarrow \sum\limits_{i=l+1}^{l+m}  {\kappa_{i} \delta}  
+\sum\limits_{i=l+1}^{l+m-1}  \left[\tan^{-1} \left(\frac{P_{i+1}} {\kappa 
_{i+1}} \right)-\tan^{-1} \left(\frac{P_{i+1}} {\kappa_{i}} \right) \right]\\
&\quad=N\pi + \tan^{-1} \left(\frac{P_{l+m+1}} {\kappa_{l+m}} \right)-\tan 
^{-1} {\left(\frac{P_{l+1}} {\kappa_{l+1}} \right)}  
\end{aligned}
\end{equation}
The exact quantization condition emerges as a limit of equation~\eqref{eq10} as 
$\delta \to 0$. In the event of $\delta \to 0$, the continuous potential 
variation is recovered, with $P_{l+m+1} \to P(\xi_{R})$ 
and $P_{l+1} \to P(\xi_{L})$. In the appendix we show that 
$P(\xi_{L})<0<P(\xi_{R})$ and 
$\left| P\left(\xi_{L,R}\right) \right|<\infty$. Further,
$\kappa_{l+1} \to \sqrt{\varepsilon-V_{q} (\xi_{L})}  = \kappa 
_{l+m} \to \sqrt {\varepsilon-V_{q} (\xi_{R})}  =0$ which, 
gives the half phase losses at the classical turning points as
\begin{equation}\label{eq11}
\lim_{\delta\to 0}
 \tan^{-1} \left(\frac{P_{l+m+1}} {\kappa_{l+m}} \right)=-\lim_{\delta\to 0}
\tan^{-1} \left(\frac{P_{l+1}} {\kappa_{l+1}} \right)=\frac{\pi} {2}. 
\end{equation}
Further,
\begin{align}
\Delta \phi_{i}&\defeq
\tan^{-1} \left(\frac{P_{i+1}} {\kappa_{i+1}} \right)
-\tan^{-1} \left(\frac{P_{i+1}} {\kappa_{i}} \right)
\nonumber\\&~=\tan^{-1} \left(\frac{P_{i+1} (\kappa_{i}-\kappa_{i+1})}
{P_{i+1}^{2} +\kappa_{i} \kappa_{i+1}} \right)\nonumber\\ 
&~=-\frac{P_{i+1} (\kappa_{i+1}-\kappa_{i})} 
{P_{i+1}^{2} +\kappa_{i} \kappa_{i+1}} 
+\mathcal{O}\left((\kappa_{i+1}-\kappa_{i})^{3}\right)  \label{eq12} 
\end{align}
which, results from expanding the inverse tangent in a tailor series in 
powers of $\kappa_{i+1}-\kappa_{i}$. Building on equation~\eqref{eq12} we obtain
\begin{widetext}
\begin{equation}
\lim_{\delta\to 0}\sum\limits_{i=l+1}^{l+m-1}  {\Delta \phi_{i}}  
=-\lim_{\delta\to 0}\sum\limits_{i=l+1}^{l+m-1}  
\frac{P_{i+1} (\kappa_{i+1}-\kappa_{i})} {P_{i+1}^{2} +\kappa 
_{i} \kappa_{i+1}}+\lim_{\delta\to 0}\sum\limits_{i=l+1}^{l+m-1}  \mathcal{O}\left((\kappa 
_{i+1}-\kappa_{i})^{3}\right)=-\int_{\xi_{L}}^{\xi_{R}}  {\frac{P} {P^{2} +\kappa^{2}} {d} \kappa}  
=-\int_{\xi_{L}}^{\xi_{R}}  {\frac{P} {\dot{P}} \dot{\kappa} {d} \xi}  
\label{eq13} 
\end{equation}
using equation~\eqref{eq9}. Thus, in the limit of $\delta \to 0$, 
equation~\eqref{eq10} reads
\begin{equation}\label{eq14}
\int_{\xi_{L}}^{\xi_{R}} \kappa - \frac{P}{\dot{P}} \dot{\kappa} \rmd  \xi
=(N+1)\pi,\quad N=1,2, \ldots,  
\end{equation}
\end{widetext}
which, is an exact criteria for the bound state wave function (specified 
through $P$) and the corresponding energy that appears in $P$ and $\kappa$. 
Interestingly, the above integral can be explicitly evaluated in terms of 
the function
\begin{equation}\label{eq15}
 Q\defeq\frac{P} {\kappa}  
\end{equation}
leading to
\begin{equation}\label{eq16}
\begin{aligned}
\int_{\xi_{L}}^{\xi_{R}} \frac{\rmd Q}{1+Q^2}
&=\tan^{-1}\left(Q(\xi_R)\right) - \tan^{-1}\left(Q(\xi_L)\right)\\
&=(N+1)\pi,\quad N=1,2, \ldots,  
\end{aligned}
\end{equation}

\section{Examples} \label{sec3}

\subsection{Symmetric power law fields} \label{sec3.1}

\begin{figure}
\includegraphics[width=\columnwidth]{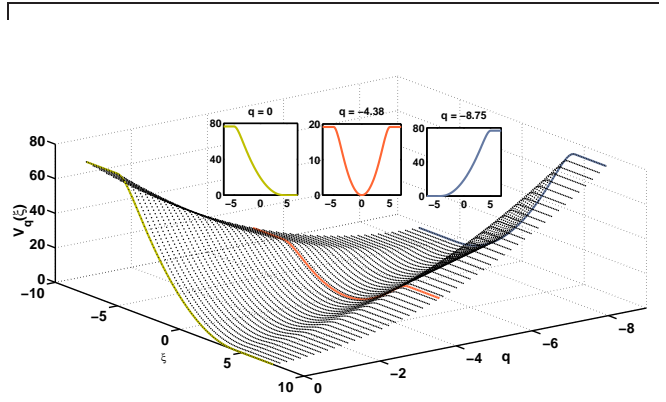}
\caption{Surface plot of $V_{q} (\xi)$ for $\beta =5,\, 
\lambda =8$. In this example, wells appear when $-8.75<q<0$. (Insets) give 
the shapes of the highlighted regions where, the red curve (middle) 
corresponds to a symmetric well considered in equation~\eqref{eq20}.
\label{fig2}}
\end{figure}

A magnetic field that varies as a simple power law is obtained by choosing
\begin{equation}\label{eq17}
B(\zeta)=\frac{1-|\zeta|^{\lambda-1}} {2} 
\left({\sgn} (\zeta +1)-{\sgn} (\zeta-1) \right)
\end{equation}
We let $\lambda >1$ to avoid a singularity at $\zeta 
=0$. In the limiting event of $\lambda \to \infty$ we approach a constant 
magnetic field. From equation~\eqref{eq2} we obtain
\begin{equation}\label{eq18}
\Phi (\zeta)=\begin{cases} 
2-\frac{2}{\lambda}, & \zeta \ge 1\\
\zeta +1-\frac{1} {\lambda} \left(1+{\sgn} (\zeta) 
|\zeta|^{\lambda}\right), &-1\le \zeta <1\\
0, &\zeta <-1
\end{cases}  
\end{equation}
Substituting $\Phi$ in equation~\eqref{eq4}, we arrive at the 
effective magnetic potential 
\begin{equation}\label{eq19}
\begin{aligned}
V_{q} (\xi)=\left(q+\xi +\beta-\lambda^{-1} \left(\frac{{\sgn} (\xi)} {\beta^{\lambda-1}} 
|\xi|^{\lambda} +\beta \right) \right)^{2}. 
\end{aligned}
\end{equation}
In this case only two 
classical turning points exist for any combination of parameters. Since 
$\Phi (\zeta)$ is an increasing curve, a well appears 
in $V_{q} (\xi)$ for $-2\beta \left(1-\frac{1} {\lambda}\right)<q<0$ 
(See Fig.~\ref{fig2}). Consequently, wave functions with $q>0$ can only scatter 
through the magnetic field region.

Hence, we look for bound states in this range. Also (as noted before) the 
Landau levels (LLs) must populate the interval $ 0<\varepsilon 
<\min \left\{q^{2},\left(q+2\beta \left(1-\lambda^{-1}\right) 
\right)^{2}\right\} $.

\begin{figure}
\includegraphics[width=\columnwidth]{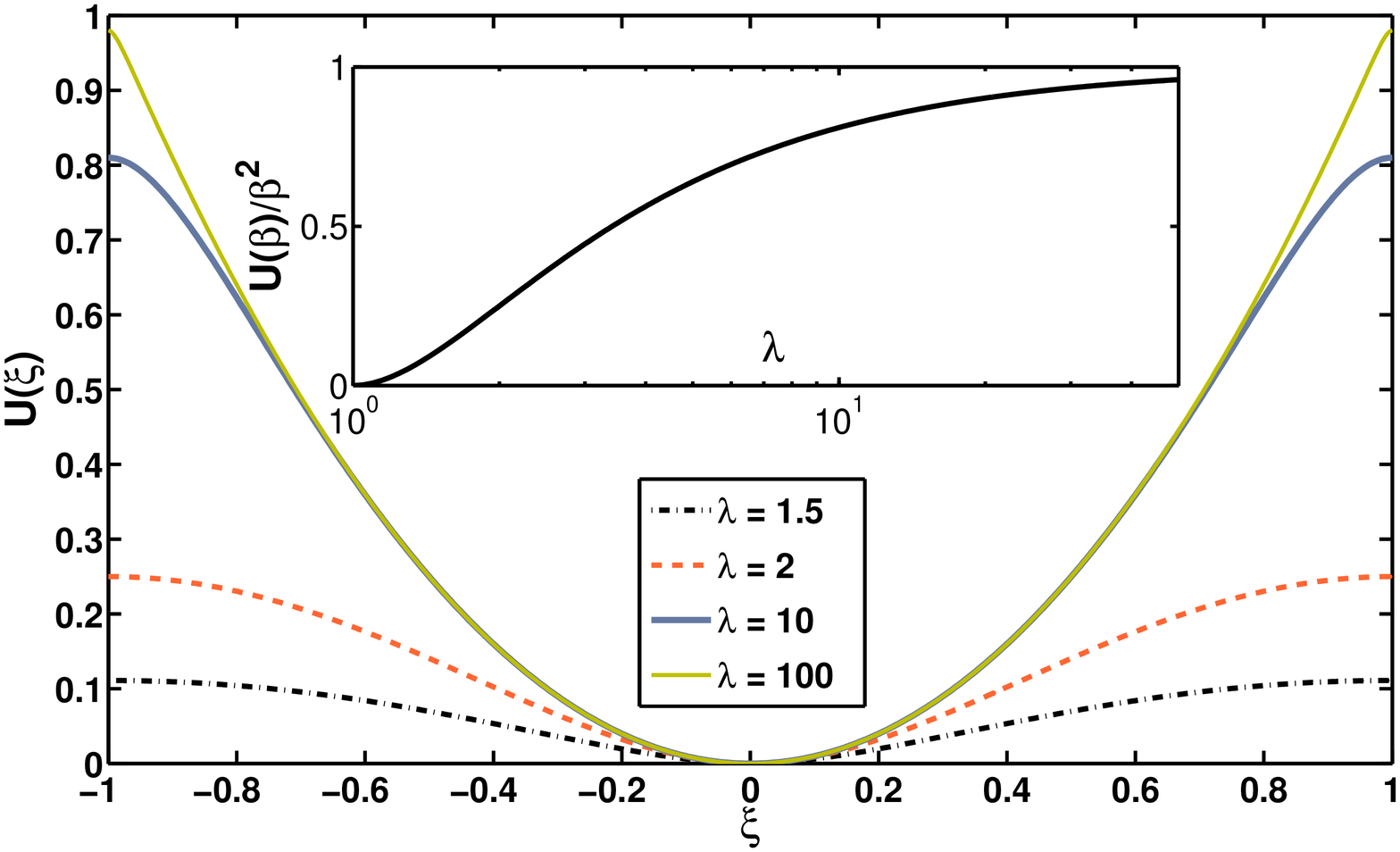}
\caption{Plots of the effective magnetic potential $U(\xi) 
(=V_{(1/\lambda-1)} (\xi))$ in the region $-1<\xi <1$, $\beta =1$. 
(Inset) gives the asymptotic levelling of the 
well-depth in the event of $\lambda \to \infty$. \label{fig3}} 
\end{figure}

For $q=-\beta \left(1-\frac{1} {\lambda}\right)$, 
the well becomes symmetric about the origin, described by
\begin{equation}\label{eq20}
U(\xi)\defeq 
V_{\beta (1/\lambda-1)} (\xi)=\left(\xi 
-\frac{{\sgn} (\xi)} {\lambda \beta^{\lambda-1}} |\xi|^{\lambda}\right)^{2}~
  |\xi|<\beta. 
\end{equation}
We plot this symmetric-potential-well for select values of $\lambda$ in 
Fig.~\ref{fig3}, with an inset displaying the variation of the well depth--given by $U(\beta)=\beta^{2} \left(1-1/\lambda \right)^{2}$--with $\lambda$. 
Note that the wells take a parabolic shape as $ \lambda \to \infty$, which 
is the case with a constant magnetic field. Also, as $ \beta \to \infty$, the effective well spans the entire axis becoming infinitely deep. Consequently, only bound states are permitted (as Landau had shown few decades ago).

\begin{figure}
\subfigure[\label{fig4a}]{\includegraphics[width=\columnwidth]{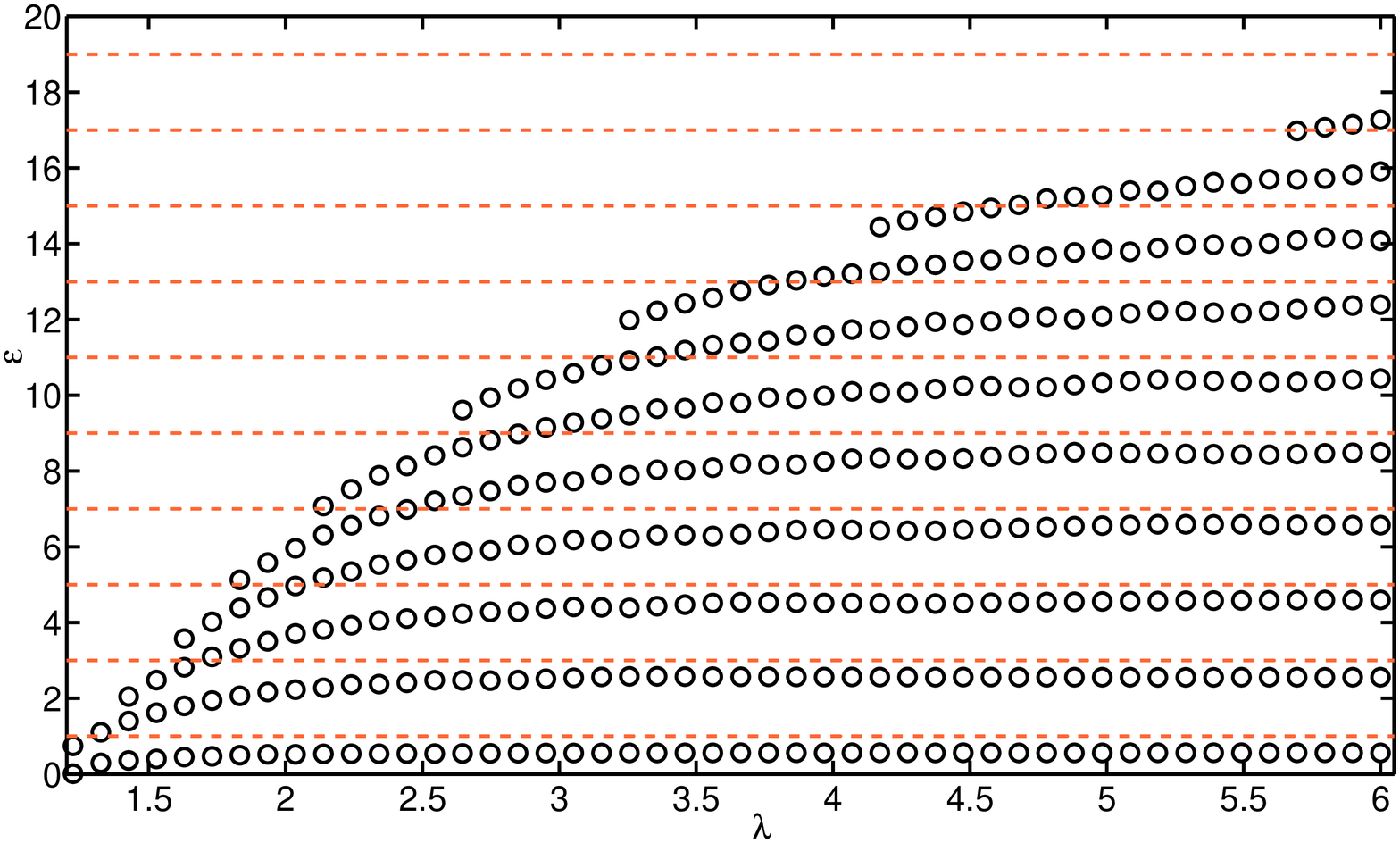}}
\subfigure[\label{fig4b}]{\includegraphics[width=\columnwidth]{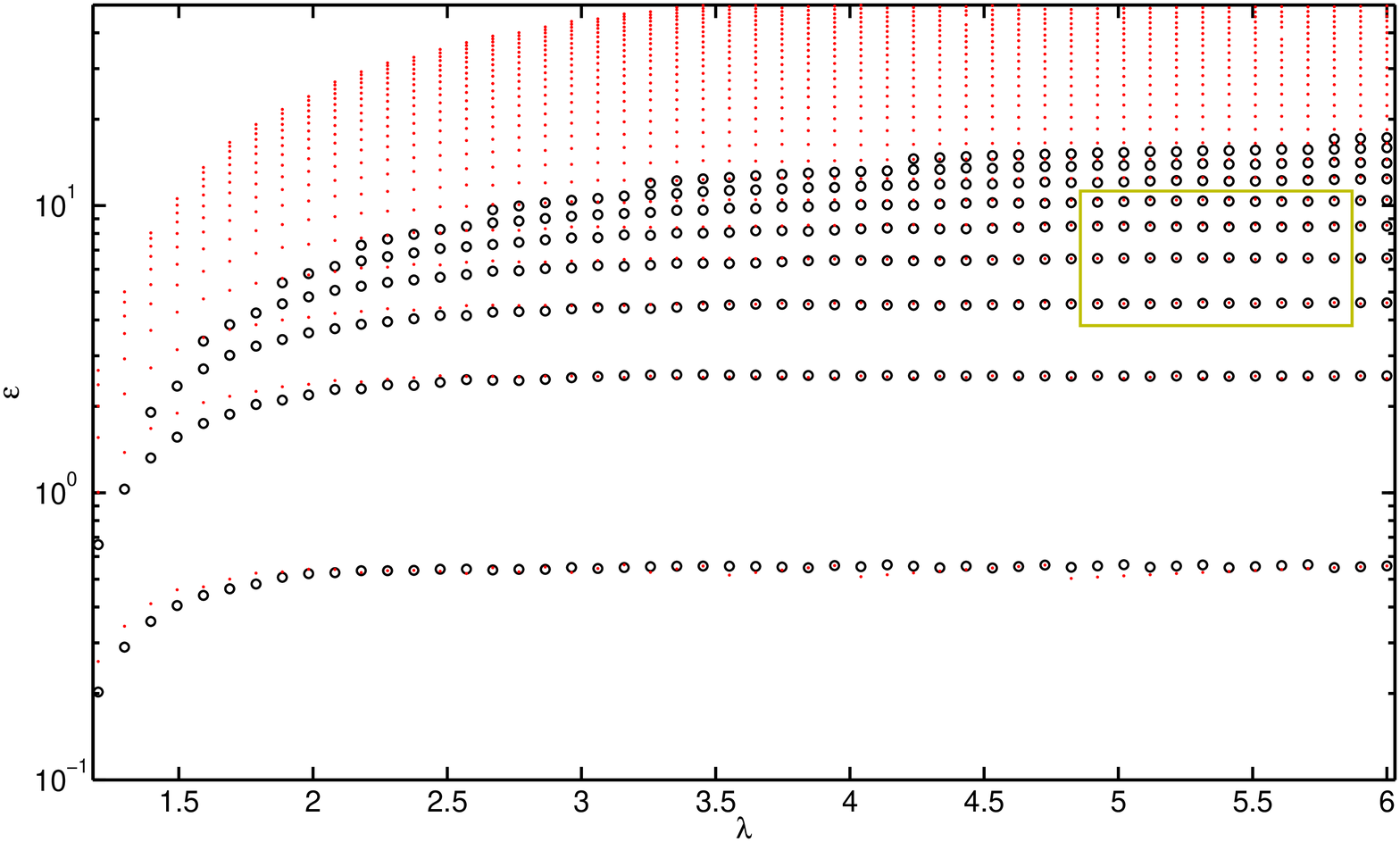}}
\caption{Plot of first few LLs (black circles) corresponding to 
different $\lambda$ for $\beta =5$. Horizontal broken lines give the 
LLs for a uniform, infinitely extending magnetic field ($LL_{l+1} =2l+1, 
l=0,1,2\ldots)$ (b) Black circles are same as that in (a), while red 
dots give the LLs for $\beta =10$.\label{fig4}} 
\end{figure}

From the bound state criterion derived above, we compute the first few LLs 
in this symmetric well in Fig.~\ref{fig4a} and study their variation with 
$\lambda$. Clearly, the LLs asymptote to those of the limiting constant magnetic field which, are shown by means of broken lines in Fig.~\ref{fig4a}---with increasing $\lambda$. Higher 
levels appear progressively, since the well-depth increases with $ \lambda$ 
(Fig.~\ref{fig3} inset). We find that the highest LL$\,\sim \beta^{2} 
\left(1-1/\lambda \right)^{2} $. The well is always brim full! Combining this observation with the asymptotic behavior of the levels noted above 
gives an adequate estimate of the total number of LLs (say $N$) populating 
a particular well is obtained (especially as $ \lambda\to\infty$).

\begin{equation}\label{eq21}
2N\sim \beta^{2} \left(1-\frac{1} {\lambda}\right)^{2} +1. 
\end{equation}
In Fig.~\ref{fig4b} we superpose the LLS of the previous case with those corresponding to a larger $\beta=10$. Numerous higher levels (red dots) appear due to the increase in the well-depth. Further, with increasing $\lambda$, the LLs for either values of $\beta$ almost overlap which is emphasized by the boxed region in Fig.~\ref{fig4b}.

Note that, even in the limit of $\lambda \to \infty$ (describing a constant magnetic field) the width of the strip remains finite (since $\beta$ is fixed); quite insensitive to which, the LLs asymptote to the LLs of a spatially-unbounded uniform field (the Landau Problem). We recall that in the former case the wave functions outside the strip decay exponentially (since the effective potential is constant out side the strip) while, those in the 
latter case have Gaussian tails. It thus turns out, despite a finite width, 
the LLs overwhelmingly approach those of the spatially-unbounded uniform magnetic field for large enough values of $\lambda$.

\subsection{2DEG under a ferromagnetic film} \label{sec3.2}

The symmetric power law magnetic fields considered above cannot be realized 
with the experimental methods discussed in Section~\ref{sec1}. However, they 
serve as good approximations to realistic magnetic fields. A crucial element 
that lacks in these ideal geometries is the presence of a fringing field outside 
the strip. For large sample sizes the fringing field can often be neglected. 
In this spirit, we had required that the magnetic field be strictly zero outside the strip (equation~\eqref{eq1}). Consequently, $V_{q} (\xi)$ became 
constant for $|\xi|>\beta$, and an exact criterion for the LLs could be obtained.

Now, we relax this constraint and allow the field to out-flank the strip; 
with an understanding that the field vanishes progressively with distance 
from the strip. For these fields, most of the treatment remains same as 
before, except for a truncation of the effective potential far away from the 
classical turning points, which turns out to be a valid approximation, if 
the LL under consideration is much below the height of the effective 
potential at the truncation points.\cite{11,12} The consequent error can be 
overcome by choosing truncation points sufficiently far from the turning 
points, allowing the LLs of interest to settle within the desired precision.

Consider a 2DEG placed under a ferromagnetic film at a distance $z$ below 
it. For a vertically magnetized film of width (thickness) $d(a)$ and 
magnetization per unit width $\sigma$, the magnetic field on the 2DEG is 
given by
\begin{gather}\label{eq22}
%\begin{aligned}
\mathbf{B}(x,z)=B_{o}B(\zeta)\mathbf{e}_{z},\nonumber \\B(\zeta)
=2\left\{\frac{(\zeta +1)} {(\zeta +1)^{2} +z_{o}^{2}}
-\frac{(\zeta-1)} {(\zeta-1)^{2} +z_{o}^{2}} \right\},\nonumber\\ 
B_{o}\defeq 2a\sigma,~z_{o} =2\frac{z} {d}  
%\end{aligned}
\end{gather}
when, $a/d, a/z\ll 1$.\cite{2} The length of the strip is infinite as 
before. Using equation~\eqref{eq2} we obtain
\begin{equation}\label{eq23}
\Phi(\zeta)=\ln \left(\frac{(\zeta +1)^{2} +z_{o}^{2}} 
{(\zeta-1)^{2} +z_{o}^{2}} \right) 
\end{equation}

\begin{figure} 
\includegraphics[width=\columnwidth]{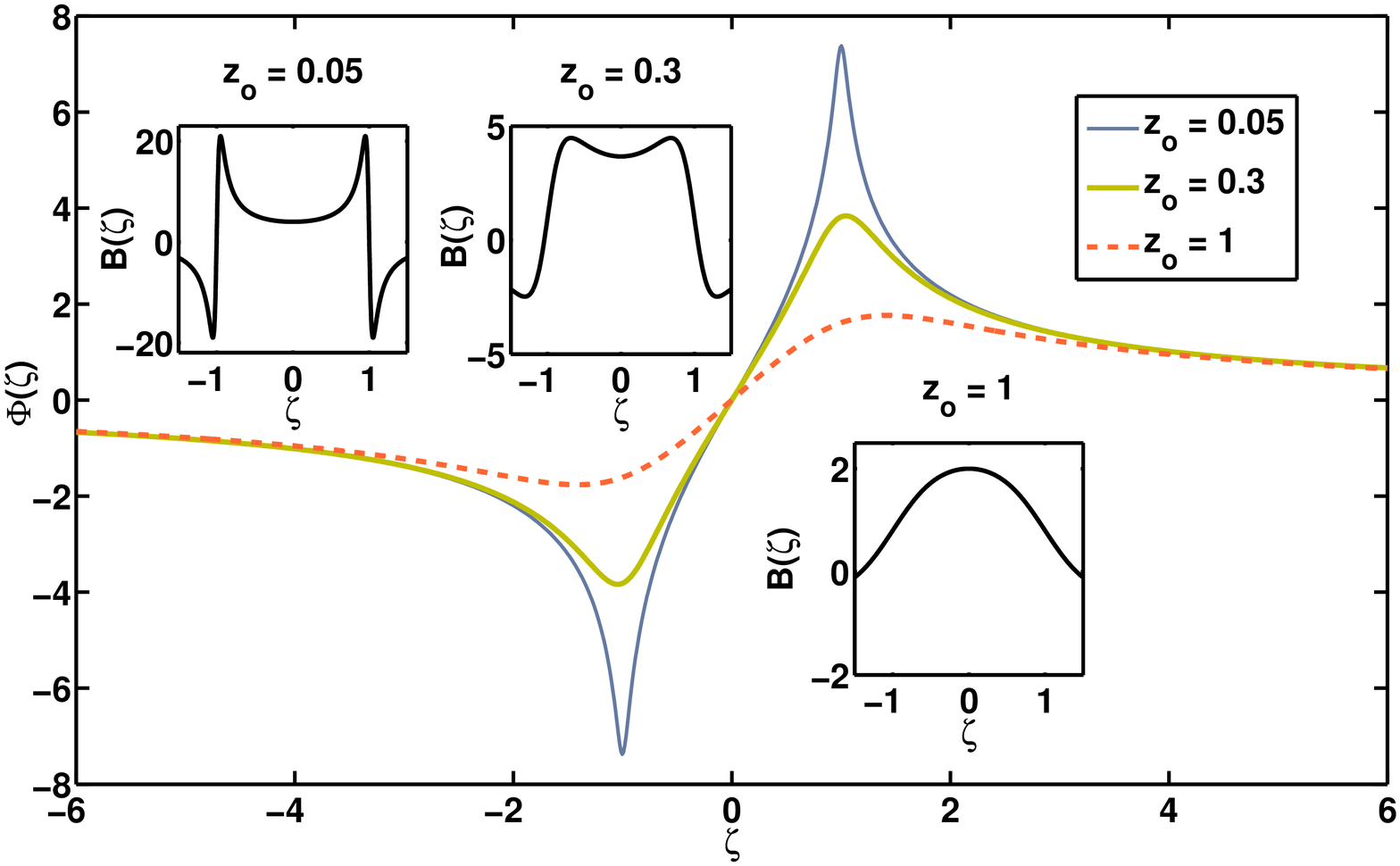}
\caption{Plot of $\Phi (\zeta)$ vs. $\zeta$ for $z_{o} =0.05,~0.3,~1$. 
(Insets) show the magnetic field variation $B(\zeta)$ for these values of $z_{o}$.
\label{fig5}} 
\end{figure} 
which, leads to the effective magnetic potential
\begin{equation}\label{eq24}
 V_{q} (\xi)=\left\{q+\beta \ln \left(\frac{(\xi +\beta)^{2} 
+\theta^{2}} {(\xi-\beta)^{2} +\theta^{2}} \right) \right\}^{2}  
\end{equation}
In obtaining the effective potential we scaled $z=\ell_{B} \theta$ and used the definition $z_{o}=2z/d=(2l_{B} /d)\theta =\theta /\beta$.

\begin{figure}
\includegraphics[width=\columnwidth]{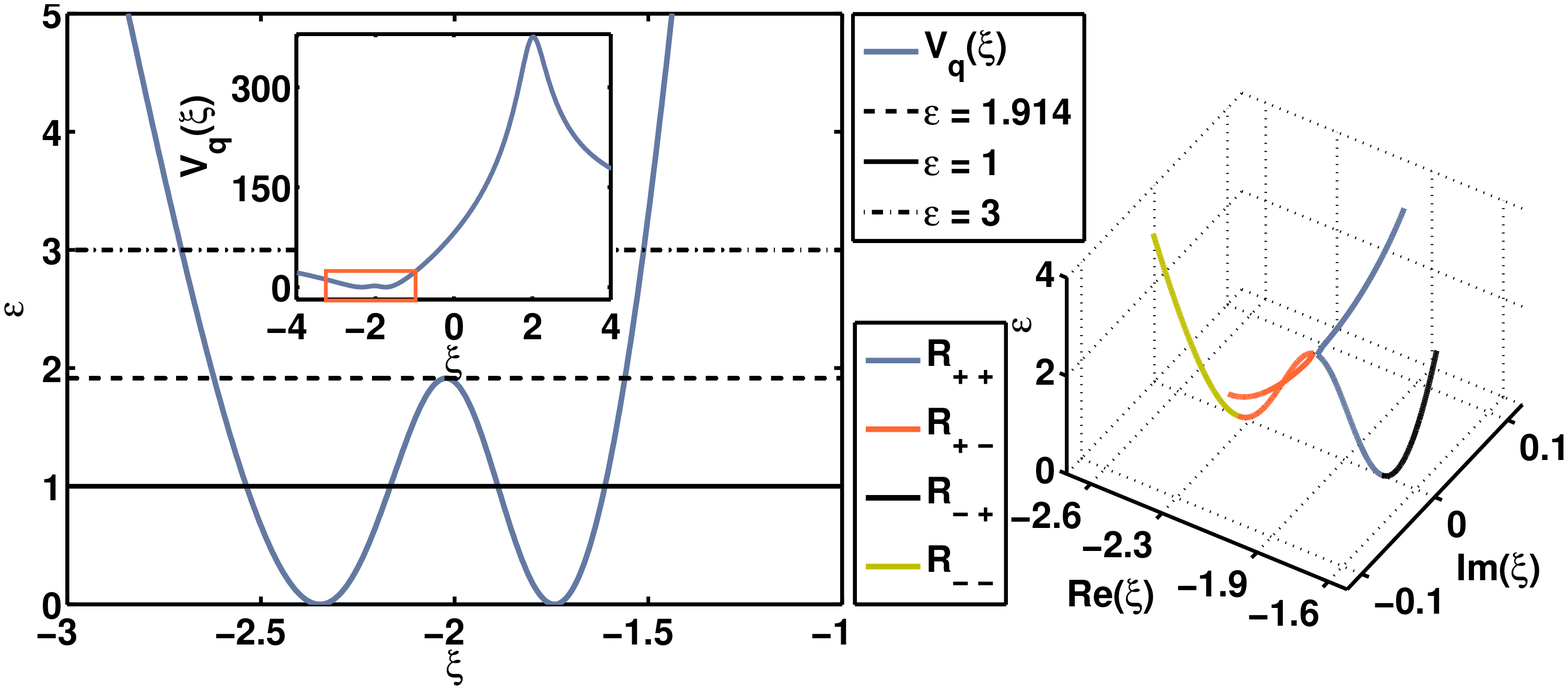}
\caption{(a) Plot of $V_{q} (\xi)$ vs. $\xi$ for $\beta =2,~\theta =0.3,~q=9$ showing a double-well potential. Horizontal energy 
lines intersect the potential at the (real) classical turning points. 
(Inset) displays the same potential on a larger range of $\xi$. (b) Loci 
of the classical turning points $R_{ab} =-\beta \coth \phi_{a} +b\left(\beta 
^{2} {\csch}^{2} \phi_{a}-\theta^{2}\right)^{1/2},~ab= \pm$ (Ref. 
equation~\eqref{eq24}) parameterized by $\varepsilon$ taken along an axis 
perpendicular to the complex $\xi$ plane. \label{fig6}} 
\end{figure}

Due to the interplay of the parameters describing the effective-potential, 
many interesting possibilities arise. First of all, unlike in the former 
example, $V_{q} (\xi)$ varies (appreciably) over the entire $\xi$ 
axis tending to $ q^{2} $ as $ |\xi|\to \infty$. 
Secondly, the effective potential in this case possesses a special reflection 
symmetry $V_{-q} (\xi)=V_{q} (-\xi)$, which 
implies that the energy eigenvalues of equation~\eqref{eq4} for 
bound state solutions remain invariant under the transformation $q\mapsto 
-q$. Thus, the LLs are \textit{doubly}  degenerate. From this property, it 
suffices to study the spectrum for $q>0$. Further, for an energy $ \varepsilon$, 
there can exist at the most four classical turning points given by
\begin{equation}\label{eq25}
\xi =-\beta \coth \phi_{\pm} \pm 
\left(\beta^{2} {\csch}^{2} \phi_{\pm}-\theta^{2}\right)^{\frac{1} {2}}~~\phi_{\pm}=\frac{q\pm \sqrt \varepsilon} {2\beta}  
\end{equation}
one or more of which might vanish--in the event of $\phi_{+},\phi_{-} =\ln 
\left(\beta /\theta \pm \sqrt {1+(\beta /\theta)^{2}} \right)$--get repeated or become complex. In the interest of bound states, four (distinct) turning points may correspond to an energy within a double-well shaped potential, while two repeated (coincident) turning points would occur when the energy hits the top of the barrier between the two wells. And, higher energies would give rise to (only) two real turning points. These cases are illustrated in Fig.~\ref{fig6}(a). A clearer perspective of the `motion' of the tuning points (in the complex $\xi$ plane) is obtained by examining their loci (Ref. Fig.~\ref{fig6}(b)) parameterized by the energy.

Next, we discuss the LLs supported by the effective potential. At the 
present moment the correct generalization of the ATMM criterion for more 
than two classical turning points is not clear, which prevents us from 
obtaining the LLs (lying below the barrier top) for the double well shown in 
Fig.~\ref{fig6}(a). However, the interested reader is referred to the work of 
L. V. Chebotarev on the `Extensions of the Bohr--Sommerfeld formula to double-
well potentials'\cite{15}  which, can be used to find the (approximate) LLs for this case.

\begin{figure} 
\subfigure[Effective magnetic potential $V_{q} (\xi)$ for $\beta =2$, 
$\theta =4$, $q=3$ with 13 LLs depicted by horizontal lines.\label{fig7a}]
{\includegraphics[width=\columnwidth]{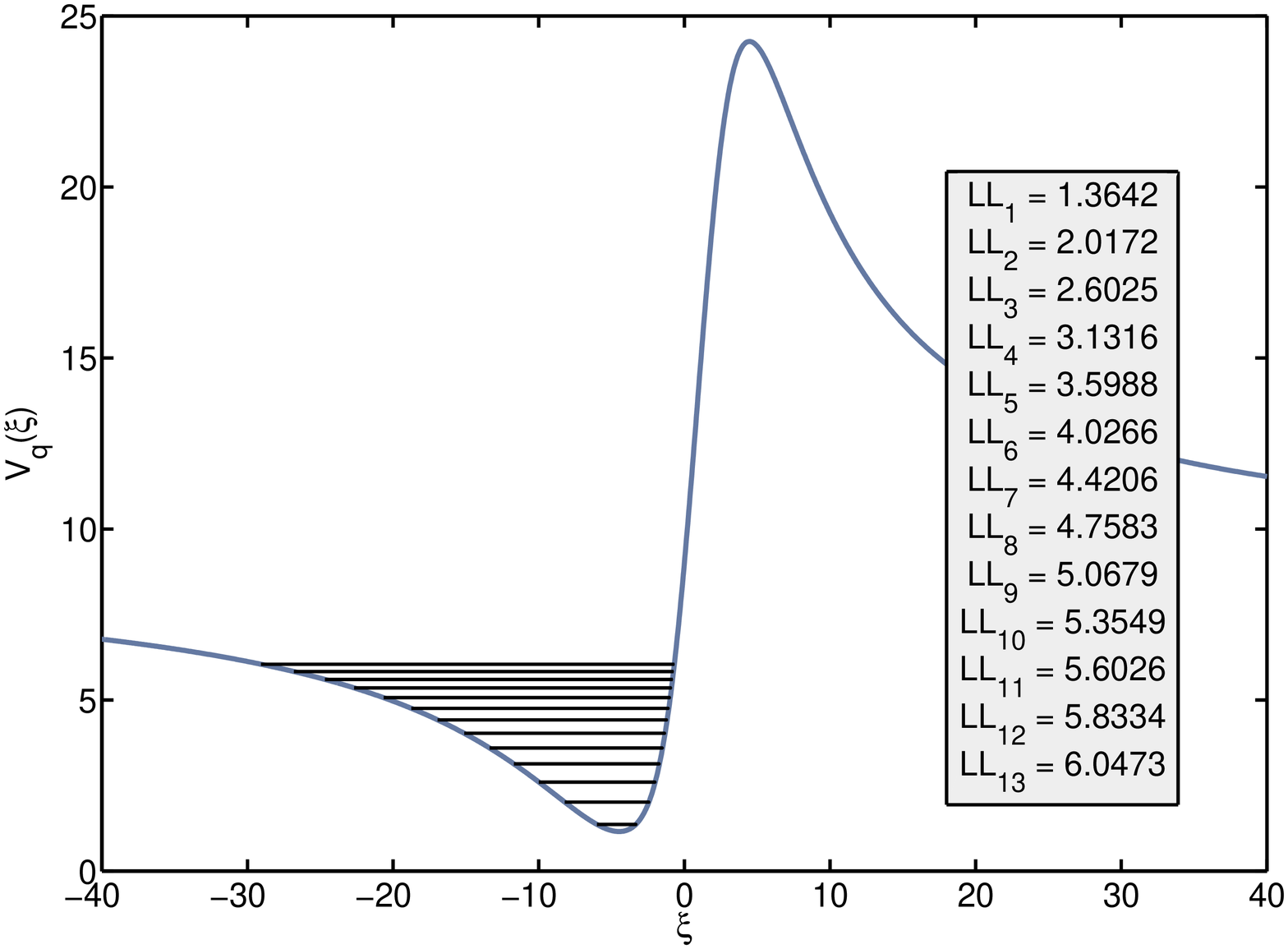}}
\subfigure[Plot of allowed LLs with $\beta =2$, $\theta =4$ and $q>2\beta \tanh^{-1} 
(\beta /\theta)\sim 2.197$. (Inset) Plot of {\#}  of LLs and the well depth 
$=q^{2}-\min (V_{q})$ vs. $q$.\label{fig7b}]
{\includegraphics[width=\columnwidth]{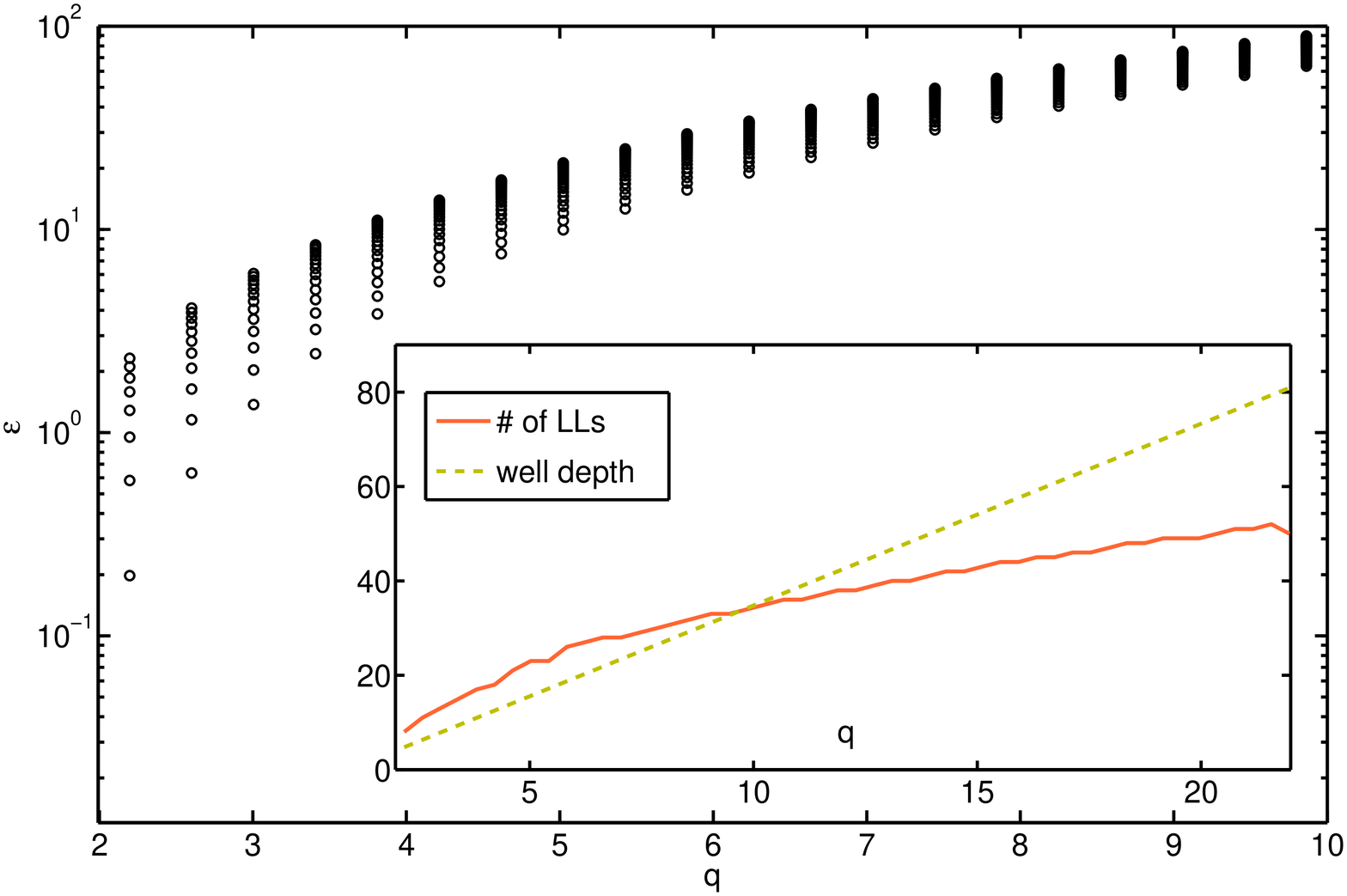}}
\caption{\label{fig7}} 
\end{figure}

In the event of $\theta >\beta,~q>2\beta \tanh^{-1} (\beta 
/\theta)$ the effective potential (in this case) offers a single well 
(Ref. Fig.~\ref{fig7a}) with two classical turning points at $\xi =-\beta 
\coth \phi_{-} \pm \left(\beta^{2} {\csch}^{2} \phi_{-}-\theta^{2}\right)^{1/2}$. The LLs in this case can be computed with our ATM quantization criterion. Using equation~\eqref{eq4} we obtain the LLs for this single well which, are plotted in Fig.~\ref{fig7b} for various values of $q$. Note that as $q$ increases, the minimum value (bottom) of the effective potential $\min (V_{q})$, also increases, hence the lowest 
Landau levels increase in energy. In fact, the well depth, i.e. $q^{2}-\min 
(V_{q})$ also increases with $q$, unfolding higher LLs. However, the rate of emergence of new LLs becomes slower with increasing $q$ as shown in Fig.~\ref{fig7b} inset. 

Before concluding, we discuss the effect of the fringing magnetic field 
pervading the region outside the strip, i.e.~$|\xi|>\beta$. 
Generally, with increasing distance from the ferromagnetic film, i.e. $\theta \gg \beta$ the fringing field can be neglected. 
However, the fringing field itself gave rise to many interesting effective 
potential shapes (unlike in the previous example). Particularly, in the 
preceding single well case with $ \theta =2\beta$, we find that the 
effective well manifests in the interval $-5\beta <\xi <0$, which lies 
outside the strip. This also gives a clue as to where the probability 
density of the electron is likely to be accumulated. 

\section{Conclusion} \label{sec4}

In this paper we considered the problem of finding the bound state solutions 
of an infinite 2DEG subjected to a perpendicular magnetic field that varies 
(arbitrarily) in one direction only. An exact criterion for the bound state 
energies or Landau levels was developed using the analytic transfer matrix 
method (ATMM) for the case when the effective magnetic potential allowed two 
(real) classical turning points. The extensions of the ATMM for more than 
two turning points is not clear at the present moment and calls for further 
consideration. Applying our formalism to a symmetric power law magnetic 
field led to the exact LLs, whose variation with the strip-width $\beta$ 
and field exponent $\lambda$ were studied. In the sequel we looked at a 
2DEG placed under a ferromagnetic film, which is an experimentally 
realizable system. Fortunately, this example could be tracked analytically 
to a great extent and the LLs for a single (effective) well were obtained 
for various values of the $y$-momentum $(\hbar/\ell_B)q$. This example also 
emphasized the role of the fringing field on the Landau levels.

\textbf{Acknowledgements:} I gratefully acknowledge the help of Prof. Dr. Hemalatha Thiagarajan whose, meticulous proofreading drew my attention to many errors which, I believe have been corrected. I also benefited from the discussions with Dr.~Sumiran Pujari on the ATM quantization formula.

\appendix*
\section{}
\setcounter{equation}{0}

Let $\psi (x)$ be a bound state solution of the one-dimensional 
Schr\"{o}dinger equation
\begin{equation}\label{eqA.1}
\frac{\rmd^{2} \psi} {\rmd x^{2}} +(E-V(x))\psi =0. 
\end{equation}
Based on the properties of an admissible bound state wave function we deduce 
an important property of the auxiliary function
\begin{equation}\label{eqA.2}
 P(x)=-\psi^{-1} \frac{\rmd \psi} {\rmd x}  
\end{equation}
which, is well defined (and bounded) at any finite $x$ excepting the nodes 
of the wave function. We show that 
\begin{equation}\label{eqA.3}
P(x)<(>)0,\qquad x=x_{L} \left(x_{R}\right) 
\end{equation}
at the left (right) \textit{most}  classical turning point $x_{L}(x_{R})$ 
that solves $V(x)=E$. Using this result we obtain the half 
phase loss at the turning point $x_{R} (x_{L})$ to be
$+\left(-\right)\frac{\pi} {2} $. It is implicit that we are working with a 
real $\psi$, hence the inequality in proposition~\eqref{eqA.3} is valid. 

Proof: The points where $V(x)< (>)E$ constitute the 
classically allowed (forbidden) region. Consider the following properties
\begin{enumerate} 
\item\label{prop1} $|\psi|\to 0, |\psi'|\to 0$ as $|x|\to \infty$
\item\label{prop2} $\psi \ne 0$ for any $x$ (no nodes) in the classically 
forbidden region
\end{enumerate} 
which, hold good for any bound state wave function. Since, equation~\eqref{eq1} 
is form invariant under the transformation $x\mapsto-x (\Rightarrow 
x_{L} \mapsto x_{R})$ while $\rmd/\rmd(-x) \mapsto 
-\rmd /\rmd(-x)$, it suffices to prove any one of 
the two propositions in~\eqref{eqA.3}. We focus on the left most classical 
turning point $x_{L} $. The truth of proposition~\eqref{eqA.3} at this point 
rejects the possibility $\sgn\left[\psi (x_{L}) 
\right]=-{\sgn} \left[\psi'(x_{L}) \right]$;
$\sgn [~]$ being the signum function. To prove this we let 
$\psi(x_{L})>0>\psi'(x_{L})$. From property~\ref{prop2}, it follows that 
$\psi (x)>0$ for all $x<x_{L}$ (a classically forbidden 
region). Therefore, $\psi^{''} >0$ (from equation~\eqref{eqA.1}--\eqref{eqA.2}). 
Since, $\psi$ is increasing (away from the origin) at $x_{L}$, it must attain 
at least one maxima before it asymptotes to the real axis (as $ x\to-\infty)$ 
remaining positive-definite all along. Clearly, at the site of this maximum 
$\psi^{''} <0$ which is not possible. The other possibility 
$\psi (x_{L})<0<\psi'(x_{L})$ (a `reflection' of the previous 
case) is readily contradicted from form-invariance of 
equations~\eqref{eqA.1}--\eqref{eqA.2} 
under the transformation $\psi \mapsto -\psi$.
 
Finally, we show that 
\begin{equation}\label{eqA.4}
| P (x_{L,R})|<\infty 
\end{equation}

Proof: Since, $\psi$ and $\psi'$ cannot vanish simultaneously,
\cite{note2} we 
need only show that a classical turning point cannot be a node of the 
wave function. Consider $x_{L} $ as before. Assume $\psi'(x_{L})\ne \psi (x_{L})=0$. 
From property~\ref{prop1}, $\psi \to 0$ 
as $ x\to-\infty$. Thus, $\psi$ must admit at least one minima (maxima) 
if $\psi'(x_{L})<0(>0)$ remaining negative (positive) definite all along. However, this contradicts the fact that 
$-\infty <x<x_{L}$ is classically forbidden.


\begin{thebibliography}{20}
\bibitem{note2}
For a bound state wave function $\psi$, if $\psi (x)=\psi'(x)=0$ at any finite
$x$ (with $|V(x)|<\infty)$, then $\psi$ must vanish identically.
\end{thebibliography}
\end{document}